\documentclass{PoS}

\title{A new concept of y-ray telescope. \\
LArGO: Liquid Argon Gamma-ray Observatory}

\ShortTitle{A new concept of y-ray telescope. LArGO: Liquid Argon Gamma-ray Observatory}

\author{\speaker{Giuseppe Andrea Caliandro}\\
        W. W. Hansen Experimental Physics Laboratory, Kavli Institute for Particle Astrophysics and Cosmology, Department of Physics and SLAC National Accelerator Laboratory, Stanford University, US. 
        Consorzio Interuniversitario per la Fisica Spaziale (CIFS), Italy\\
        E-mail: \email{caliandr@slac.stanford.edu}}

 \author{Biagio Rossi   \\
         Princeton University, US.
         Universit\`{a} degli studi di Napoli \textquotedblleft Federico II \textquotedblright, Italy.
         Istituto Nazionale di Fisica Nucleare (INFN), sezione di Napoli, Italy}
 
\author{Francesco Longo\\
        Universit\`{a} di Trieste, Italy.
        Istituto Nazionale di Fisica Nucleare (INFN), sezione di Trieste, Italy}

\author{Giuliana Fiorillo\\
        Universit\`{a} degli studi di Napoli \textquotedblleft Federico II \textquotedblright, Italy.
        Istituto Nazionale di Fisica Nucleare (INFN), sezione di Napoli, Italy}

\author{Claudio Labanti\\
        Istituto di Astrofisica Spaziale e Fisica Cosmica (INAF), sezione di Bologna, Italy}

\author{Federico Sanchez\\
        Institut de Fisica d'Altes Energies (IFAE), Barcelona, Spain}

\author{Thorsten Lux\\
        Institut de Fisica d'Altes Energies (IFAE), Barcelona, Spain}

\abstract{
LArGO (Liquid Argon Gamma-ray Observatory) consists of a new design for a $\gamma$-ray telescope, which exploits the idea of using a Liquid Argon Time Projection Chamber (LAr-TPC) as tracker-converter. 
Particle tracking in LAr-TPC can efficiently starts since the primary photon vertex. Indeed, while in the present space telescopes the incident photon converts in a tungsten foil, which is a passive material, in a LAr-TPC this conversion happens in LAr itself, which is fully active.
In this proceeding is described a plausible design for the tracker-converter detector which fulfills the constraints on conversion efficiency, angular resolution, and wide field of view.
It is demonstrated how this design can provide an unprecedented angular resolution for a $\gamma$-ray telescope, leading to a significant improvement in sensitivity and most important disclosing the possibility to detect the polarization of $\gamma$-ray emission. 
}

\FullConference{XV International Workshop on Neutrino Telescopes\\
                 March 11-15 2013\\
                 Venice, Italy}

\begin{document}

\section{Introduction}

In recent years $\gamma$-ray astronomy is experiencing a revolution in terms of new discoveries. A huge scientific production is ongoing in this field, as well as very
important steps forward in the understanding of astrophysical sources have been already obtained. Anyway, we are still far from the complete understanding of
sources like pulsars, AGN, $\gamma$-ray bursts, $\gamma$-ray binary systems, etc.
In the GeV energy range, this era started with the launches of the new generation $\gamma$-ray satellites AGILE \cite{2003SPIE.4851.1151T} and Fermi \cite{2009ApJ...697.1071A}, in April 2007 and June 2008 respectively.

The basic design of a $\gamma$-ray telescope working in the pair conversion energy range requires three sub-detectors: an anti coincidence detector to veto charged particles, a calorimeter below the tracker measures the energy of the pair, and finally the principal one is a tracker-converter which has the double task to make the photon converts, and to track the produced pair. From this double task arises an ambiguity for the tracker converter, because a large amount of
converting material (high Z) is necessary to maximize the effective area, but at the same time it needs to be minimized to reduce the effect of the multiple scattering. 

The solution adopted by both AGILE and the Large Area Telescope (LAT) on board Fermi 
is to use thin layers of passive converting material (W) interleaved with active
layers of SSD, which record the passage of charged particles.

A similar ambiguity is present in neutrino detectors which need an extremely massive target and a good tracking resolution to observe the vertex of neutrino interactions. The solution adopted by the ICARUS collaboration \cite{icarus} is to use a liquid argon TPCs where the massive material is also active for the tracking: a full active detector without dead zones.

Motivated by this analogy, we are investigating the idea of a new design of $\gamma$-ray telescope based on the use of a liquid argon TPC as tracker-converter.
%
% Alternatively, the use of gaseous TPC as tracker-converter for a $\gamma$-ray pair telescope was first suggested
% by Hartman \cite{45} and further explored by Bloser, et al. \cite{46} \cite{47}, Hunter et al. \cite{48} \cite{49}, Ueno et al. \cite{50}, Bernard \cite{2013NIMPA.729..765B}, and Hunter et al. \cite{}.
%
In the next section we describe how we adapt the liquid argon TPC to be used as $\gamma$-ray detector.

\section{Tracker design}

We are going to design the tracker of LArGO in order that it fulfills the 
following conditions.
Our purpose is to project a $\gamma$-ray telescope which has better performance than Fermi-LAT, 
and in addition can work as $\gamma$-ray polarimeter at least until photon energies of 1 GeV.
In order for the polarization to be detected, the plane of the converted pair needs to be identified. To make it possible, the tracker has to discriminate the two tracks with an angular resolution ($\sigma_{\theta}$) lower than their aperture angle $\theta$. The most probable value of $\theta$ is \cite{1962PhRv..126..310M} 
\begin{equation}
  \theta \sim \frac{4 m_e}{E_{\gamma}},
  \label{e1}
\end{equation} 
where $m_e$ is the electron mass, and $E_{\gamma}$ is the energy of the incident photon.
Approximately, it is $\theta = 1$ deg at 100 MeV, and $\theta = 2$ mrad at 1 GeV.
The tracker of a $\gamma$-ray detector has also the task to efficiently convert the incident photons. In order for this task to be fulfilled the depth of the tracker need to be equivalent to one radiation length, as in the case of Fermi-LAT, or greater.
In contrast, to have a large field of view, and to be suitable for a satellite, its full height has to be compact of the order of one meter.

Current TPCs have typical pitch of the readout plane $l = 1$ mm, and  spatial resolution $\sigma = 0.1$ mm.
In LAr-TPC, where there is not charge amplification of the drift electrons, these characteristics are constrained by the noise of the readout front-end electronic. Current R\&D studies for new neutrino and dark matter experiments are developing electronic systems which compared with those used in ICARUS are from three times \cite{2012JInst...7C2004C} to more than 10 times \cite{2012JInst...7.5010A} less noisy. With these systems it is possible to reduce the pitch untill $l=0.1$ mm, and to improve the spatial resolution to at least a quarter of the pitch $\sigma = 0.025$ mm, when the readout is composed by three grids of wires or strips (two of induction, and one of collection).

Still, with these characteristics a tracker composed by a compact liquid argon TPC as depth as its radiation length ($X_0 = $ 14.3 cm) has not the desired angular resolution, as we can calculate using the formulas summarized in \cite{Innes}. 
%We also considered the option of gaseous TPCs, which have the requested angular resolution, but their radiation length is really too large. For example, at 1 bar the Xe gas has $X_0 \sim 50$ m. To reduce its $X_0$ to one meter, is necessary to compress it at 50 bar. Currently, studies of high pressure TPC are in developments \cite{}, but at pressures not higher than 10 bars. Then, we discarded this hypothesis.     
%
Then, the solution we adopted to satisfy all the constraints is to design a tracker composed by a column of several thin LAr TPCs separated by a certain distance. Similarly to the structure of the current $\gamma$-ray telescopes, but instead to have layers of SSDs and tungsten foils, LArGO will have layers of LAr TPCs. 

The space among the TPC-layers is necessary in order to well discriminate in a TPC-layer the two tracks of the pair converted in the layer just on top.  
Then, the spacing is determined by the pair aperture angle and the spatial resolution of the TPC. 
With a distance of $d=2.5$ cm the tracks of a pair by 1 GeV photon are separated at the second layer by twice the TPC pitch.

The thickness ($L$) of the TPC-layers has to be chosen in order that the angular deflection due to the multiple scattering affecting the tracks is lower than the desired angular resolution.
At high energies, the 68\% of the multiple scattering angle distribution is approximately
\begin{equation}
  \theta_{MS} = \left(\frac{13.6 {\rm MeV}}{E_e} \right) \sqrt{\frac{L}{X_0}}.
  \label{e2}
\end{equation} 
In the hypothesis of energy equipartition ($E_{e^+} = E_{e^-} = E_{\gamma}/2$), the thickness L is given by equating Eq. \ref{e1} and \ref{e2}, resulting in $L=0.8$ mm. This thickness is extremely small. The situation can be improved setting the pressure of the liquid argon close to its triple point (68.89 kPa). In this case $X_0 = 204$ mm, and $L = 1.1$ mm, but it is still too small because  
to reach a depth of 1 $X_0$ for the whole detector there would be 180 layers, that spaced 2.5 cm each give a total length of 4.5 m.
We decided to determine the thickness $L$ in order to satisfy the other requirements, rather than the angular resolution 
\begin{eqnarray}
 L \cdot N_{layers} = X_0 \\
 N_{layers} (L+d) < 1 \, {\rm m}.
 \label{e3}
\end{eqnarray}
The best solution we found is to have 32 TPC-layers with thickness $L=6.5$ mm at a pressure of 70 kPa.
The price to pay with this solution is that only the photons converting in the bottom 1.1 mm of each layer are suitable to detect the polarization. Approximately, they are 17\% of the statistics.

% %
% \begin{equation}
%   L = \left(\frac{13.6 {\rm MeV}}{E_{e}} \right)^2 \frac{L}{X_0}.
%   \label{e3}
% \end{equation} 
% %
% The liquid argon has at 1 bar $X_0 = 143$ mm, resulting in $L \sim 3$ mm.  

% There are two problems with the values obtained by our calculations. With $L=3$ mm there are only 3 points per track in a single TPC layer, and a depth of 1 $X_0$ for the whole detector means 47 layers, that spaced 5 cm each make a total length of 2.5 m.
% A plausible solution is to reduce the pressure of the liquid argon to half bar, and increase the thickness of the layers to $L=12$ mm. In this way the radiation length duplicate, but the number of planes is reduced by a factor of two, giving a total detector length of 1.4 m, better in agreement with the constraints. In each layer there are 12 points per track, but the price to pay is that only the photons converting at a depth greater than 6 mm in each layer are suitable to detect the polarization. Approximately, they are 50\% of the statistics.  

\begin{figure}
\includegraphics[width=0.49\textwidth]{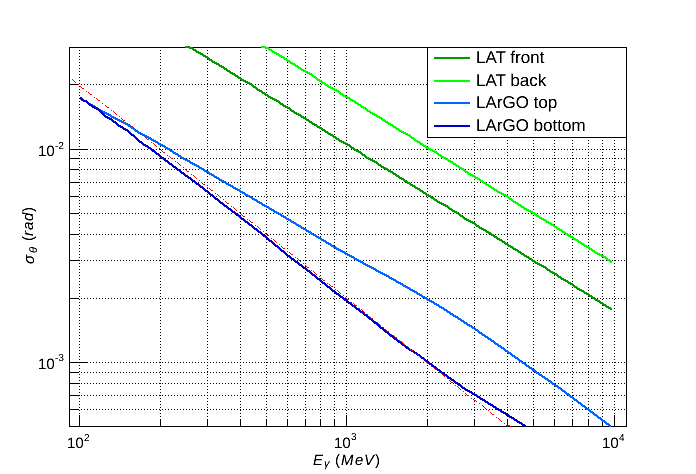}
\includegraphics[width=0.49\textwidth]{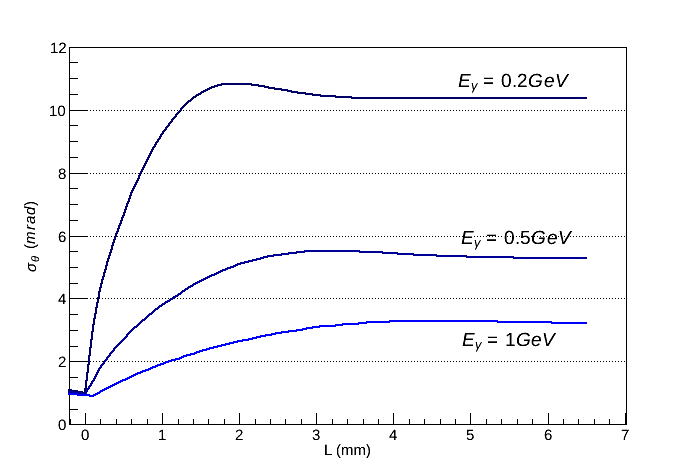}
 \caption{{\it Left.} Angular resolution of LArGO versus the energy of incident photons converting in the top of a TPC-layer (LArGO top), or at one millimeter from its bottom (LArGO bottom). For reference, the angular resolution of LAT (front/back) is plotted. The dot dashed red line is the upper limit for polarization detection.
 {\it Right.} Profile of the angular resolution versus the depth of the conversion point in a TPC-layer  for photons with energy 200 MeV, 500 MeV, and 1 GeV. $L=0$ mm is deepest point of a TPC-layer.
 \label{fig1}}
 \end{figure}

Billoir \cite{Billoir} describes a recursive procedure which backward propagates the results of the track fit from the last measured point to the first one, adding the information gain and loss at each measurement point.
This method provides a straightforward means to calculate the errors on the track parameters at its starting point.
We implemented this method to estimate the angular resolutions of the LArGO tracker.
Fig. \ref{fig1} left panel shows the angular resolution versus the energy of the photons converting at the top of a TPC-layer (light blue line), and at 1 mm from the very bottom (dark blue line). The dot dashed red line represents the upper limit of the angular resolution for the polarization detection. It is interesting to note that while at 1 GeV only the photons converted at the bottom of the layer have resolution below the polarization limit, at low energies $\lesssim 200$ MeV all the photons are useful to detect the polarization.  

In the right panel of Fig. \ref{fig1} is plotted the angular resolution versus the depth of the conversion point in a TPC-layer of photons with energy 200 MeV, 500 MeV, and 1 GeV.
We can see that starting from the bottom of a TPC-layer ($L=0$ mm) the angular resolution rises until its maximum value, and then decreases a bit until a plateau. The rise is much steeper at low energy than high energies.  
The mean angular resolution at 1 GeV is $\sim 3$ mrad, that is better than the LAT angular resolution (front + back) by a factor $\sim5$. In the hypothesis of equal effective area, LArGO would improve the sensitivity to point-like $\gamma$-ray sources by the same factor. At low energies the improvement is even greater.    

\section{Conclusions}

LArGO is a project in development which is born from the idea to adapt a technique developed for neutrino physics to the needs of $\gamma$-ray astronomy.
In this proceeding is described a new concept of tracker-converter for $\gamma$ rays based on the use of liquid argon TPCs. 
It is designed in order to obey the requisites of high conversion efficiency, large field of view, and unprecedented angular resolution which allow polarization detection.
This last point is of particular interest, because the polarization of high energy $\gamma$ rays is totally unexplored by the current telescopes, while the information it brings is of key importance for the understanding of $\gamma$-ray sources, especially those whose emission is magnetically driven, like pulsars, AGNs, micorquasars, and GRBs.

Alternatively, the use of gaseous TPC as $\gamma$-ray polarimetry  was first suggested
by Hartman \cite{45} and further explored by Bloser, et al. \cite{46}, Ueno et al. \cite{50}, Bernard \cite{2013NIMPA.729..765B}, and Hunter et al. \cite{Hunter}.
Gaseous TPCs have a straightforward angular resolution and allow a more accurate study of polarization than LArGO, but in contrast they lack in conversion efficiency. 

Fermi-LAT is the best performing $\gamma$-ray telescope currently operative on-orbit. Compared to it, LArGO has a better angular resolution and sensitivity by at least a factor $5$. In other words, LArGO has at 100 MeV approximately the same angular resolution that currently Fermi-LAT has at 1 GeV. Of course, this allows for a much higher quality mapping of the $\gamma$-ray sky.


\begin{thebibliography}{99}

\bibitem[1]{2003SPIE.4851.1151T} Tavani, M., Barbiellini, 
G., Argan, A., et al.\ 2003, {\it The AGILE instrument. X-Ray and Gamma-Ray Telescopes and 
Instruments for Astronomy. Proceedings of SPIE} Vol., {\bf 4851}, 1151

\bibitem[2]{2009ApJ...697.1071A} 
Atwood W.~B., et al., 2009, {\it The Large Area Telescope on the Fermi Gamma-Ray Space 
Telescope Mission. The Astrophysical Journal}, {\bf 697}, 1071 

\bibitem[3]{icarus} ICARUS Collaboration, 1992, {\it The ICARUS liquid argon TPC: a complete imaging device for particle physics. Nuclear Instruments and Methods in Physics Research A}, {\bf 315}, 223

\bibitem[4]{1962PhRv..126..310M}Maximon, L.C.~and 
Olsen, H., 1962, {\it Measurement of Linear Photon Polarization by Pair 
Production}. {\it Physical Review} {\bf 126}, 310.

\bibitem[5]{2012JInst...7C2004C} 
Chen H., et al., 2012, {\it Readout electronics for the MicroBooNE LAr 
TPC, with CMOS front end at 89K. Journal of Instrumentation} {\bf 7}, 2004.

\bibitem[6]{2012JInst...7.5010A} 
Auger M., et al., 2012, {\it The EXO-200 detector, part I: detector design and 
construction. Journal of Instrumentation}, {\bf 7}, 5010. 

\bibitem[7]{Innes} Innes 
W.~R., {\it Some formulas for estimating tracking errors. Nuclear Instruments and Methods 
in Physics Research A} {\bf 329}, 238-242.

\bibitem[8]{Billoir} 
Billoir P., 1984, {\it  Track 
fitting with multiple scattering: A new method. Nuclear Instruments and 
Methods in Physics Research} {\bf 225}, 352-366. 


\bibitem[9]{45} R.C. Hartman, 1989, {\it Astronomical Gamma Ray Telescopes in the Pair Production Regime. Nucl. Phys. B}, Proc. {\bf 535} Suppl. 10, 130-138

\bibitem[10]{46} P.F. Bloser, et al. 2005, {\it Gas Micro-well Track Imaging Detector for Gamma-ray
Astronomy. UV, X-Ray, and Gamma-Ray Space Instrumentation for Astronomy XIV (San Diego, CA: SPIE)}, {\bf 5898}, 152-163

%\bibitem[11]{47} P.F. Bloser, et al. 2004, {\it A Concept for a High-Energy Gamma-ray Polarimeter. SPIE} {\bf5165} 322-333

%\bibitem[48]{48} S.D. Hunter, et al. 2010, {\it Development of the Advanced Energetic Pair Telescope (AdEPT) for Medium-Energy Gamma-Ray Astronomy, SPIE Space Telescopes and Instrumentation}, {\bf 7732}:773221

%\bibitem[49]{49} S.D. Hunter, et al. 2012, {\it Development of a telescope for medium energy gamma-ray astronomy, Proceedings of SPIE} Vol. {\bf 8443}, 84430F

\bibitem[11]{50} K. Ueno, et al. 2011, {\it Development of the tracking Compton/pair-creation c
amera based on a gaseous TPC and a scintillator camera}. {\it Nuclear Instruments and Methods in Physics Research A} {\bf 628}, 158 

\bibitem[12]{2013NIMPA.729..765B}Bernard, D., 2013, {\it Polarimetry of cosmic gamma-ray sources above e$^{+}$e$^{-}$ pair creation threshold}. {\it Nuclear Instruments and Methods in Physics Research A} {\bf 729}, 765. 
  
\bibitem[13]{Hunter} 
Hunter S.~D., et al., 2013, {\it A Pair Production Telescope for Medium-Energy Gamma-Ray 
Polarimetry}. arXiv:1311.2059 



\end{thebibliography}
\end{document}